\documentstyle[prl,preprint,aps]{revtex}
\begin{document}
\draft

\title{\bf EASY MONITORED ENTANGLED STATE}

\author{M.A. Can, A.A. Klyachko, and A.S. Shumovsky}

\address{Faculty of Science, Bilkent University, Bilkent, Ankara,
06533 Turkey}

\maketitle

\begin{abstract}
We discuss the generation and monitoring of durable atomic
entangled state via Raman-type process, which can be used in the
quantum information processing.
\end{abstract}


\narrowtext


The problem of creation of entangled states in atomic systems has
attracted a great deal of interest (see$^1$ and references
therein). In particular, the entangled states were engineered
through the use of cavity QED$^2$ and technique of ion traps$^3$.

An interesting proposal has been made recently$^4$ (for further
discussion, see Refs. 5 and 6). It was shown that a pure entangled
state of two atoms in an optical resonator can be obtained through
the exchange by a single photon. Since the excitation of the
system either is carried by a cavity photon or is shared between
the atoms, the absence of the photon leakage from the resonator
can be associated with the presence of atomic entanglement. This
entanglement can be observed in the process of continuous
monitoring of the cavity decay$^4$. The importance of this scheme
is caused by the fact that its realization seems to be easy
available with present experimental technique. The result can also
be generalized on the multi-atom systems$^6$.

In view of the practical realization, it seems to be more
convenient if the existence of atomic entanglement would manifest
itself via a certain signal photon rather than via the absence of
photons as in Ref. 4. This implies that there should be at least
two different modes interacting with the atoms such that the
photon of one of them provides the correlation between the atoms,
while the photon of the other mode can freely leave the resonator
to signalize the rise of atomic entanglement.

In this note we discuss a way how to obtain a durable maximum
entangled state of atoms in an optical resonator which can be
monitored through the detection of signal photons.

Consider the Raman-type process in a three-level atom shown in
Fig. 1. Here $1 \leftrightarrow 2$ and $2 \leftrightarrow 3$ are
the dipole transitions corresponding to the pump and Stokes modes,
while the dipole transition between the levels 1 and 3 is
forbidden because of the parity conservation. We assume that the
two identical atoms of this type are located in a high-quality
cavity tuned to resonance with $1 \leftrightarrow 2$ transition,
while the Stokes photons can leak away freely (Fig. 2).

Assume that initially both atoms are in the ground state (level 1)
and there is a single cavity photon, so that the initial state is
\begin{eqnarray}
| \psi_0 \rangle = |1,1 \rangle |1_P \rangle |V_S \rangle .
\label{1}
\end{eqnarray}
Here $|n_P \rangle$ denotes the $n$-photon state of the cavity
(pump) mode and $|V_S \rangle$ denotes the vacuum state of the
Stokes field. Then, the absorption of the cavity photon by atomic
system should lead to the state
\begin{eqnarray}
| \psi_1 \rangle = \frac{1}{\sqrt{2}} (|2,1 \rangle +|1,2 \rangle
) |0_P \rangle |V_S \rangle , \label{2}
\end{eqnarray}
which manifests the entanglement of atoms excited to the level 2.
This atomic entanglement is similar to that discussed in Ref. 4
and has a very short lifetime defined by the atom-field coupling
constants for the allowed transitions. The decay of the excited
atomic state (2) can either return the system into the initial
state (1) or turn (2) into the state
\begin{eqnarray}
| \psi_k \rangle = \frac{1}{\sqrt{2}} (|3,1 \rangle +|1,3 \rangle)
|0_P \rangle |1_{Sk} \rangle , \label{3}
\end{eqnarray}
where $|n_{Sk} \rangle$ denotes the state of $n$ Stokes photons
with frequency $\omega_{Sk}$. This state again manifests the
maximum atomic entanglement. Since the cavity walls are supposed
to be transparent for the Stokes photons and $3 \leftrightarrow 1$
is the dipole-forbidden transition, the atomic entanglement
described by (3) would exist for a very long time determined by
the weak interaction between the atoms excited to the level 3 and
a certain dissipative environment. The creation of this atomic
entanglement manifests itself by the Stokes photon that can be
detected outside the cavity.

It should be noted that, in addition to $|\psi_1 \rangle$ and
$|\psi_k \rangle$, the following maximum entangled states
\begin{eqnarray}
| \phi_1 \rangle = \frac{1}{\sqrt{2}} (|2,1 \rangle -|1,2 \rangle
) |0_P \rangle |V_S \rangle , \nonumber \\ | \phi_k \rangle =
\frac{1}{\sqrt{2}} (|3,1 \rangle -|1,3 \rangle) |0_P \rangle
|1_{Sk} \rangle  \nonumber
\end{eqnarray}
also contribute into  the base states of the system under
consideration. Both of them are stabile states but they cannot be
achieved in the process of evolution beginning with the initial
state (1) (see Ref. 6). Therefore, they can be discarded.

To describe the quantum dynamics of the system, we note that the
upper atomic level $2$ can be adiabatically removed$^7$ (also see
Ref. 8 and references therein). In this case, the two-photon
transitions in effective two-level atoms described by the
effective interaction Hamiltonian
\begin{eqnarray}
H_{int}= \sum_k \sum_{f=1}^2 \lambda_k \{ R_{31}(f)a^+_{Sk}a_P+H.c
\}  \label{4}
\end{eqnarray}
should be considered. Here $\lambda_k$ denotes an effective
coupling constant has been defined in Ref. 7 and $R_{ij}(f)$ is
the atomic operator corresponding to the transition $j \rightarrow
i$ in the $f$-th atom. Under the influence of (4), the initial
state (1) is directly transformed into (3), so that the
intermediate entangled state (2) can be omitted. Then, the
time-dependent wave function of the system takes the form
\begin{eqnarray}
| \Psi (t) \rangle = C_0(t)| \psi_0 \rangle + \sum_k C_k(t)|
\psi_k \rangle , \label{5}
\end{eqnarray}
where the time-dependent coefficients are defined by the
Schr\"{o}dinger equation together with the initial condition
\begin{eqnarray}
| \Psi(0) \rangle = | \psi_0 \rangle , \quad C_0(0)=1, \quad
C_k(0)=0. \label{6}
\end{eqnarray}
Taking into account that the total Hamiltonian has the form
\begin{eqnarray}
H=H_0+H_{int}, \nonumber \\ H_0 = \omega_P a^+_Pa_P + \sum_{k}
\omega_{Sk} a^+_{Sk}a_{Sk} + \omega_{31} \sum_{f=1}^{2} R_{33}(f)
, \nonumber
\end{eqnarray}
we get the following system of linear differential equations
\begin{eqnarray}
i \dot{C}_0 & = & \omega_P C_0+ \sum_k \lambda_k \sqrt{2} C_k,
\nonumber \\ i \dot{C}_k & = & ( \omega_{Sk} + \omega_{31} )C_k+
\lambda_k \sqrt{2} C_0 . \label{7}
\end{eqnarray}
Here $\omega_{31} =E_3-E_1$ denotes the energy difference between
the levels 3 and 1 connected by the two-photon transition. These
Eqs. (7) together with the initial conditions (6) completely
determine the evolution of the state (5). Using the standard
methods$^9$, it is easy to show that the system evolves from the
initial state (1) into the final state
\begin{eqnarray}
| \Psi (t) \rangle \rightarrow \sum_k J_k | \psi_k \rangle ,
\nonumber
\end{eqnarray}
corresponding to the maximum atomic entanglement described by (3).
Here
\begin{eqnarray}
J_k = \frac{-i \lambda_k \sqrt{2}}{\gamma /2 -i( \omega_{Sk} +
\omega_{31} - \omega_P - \Delta )}, \nonumber
\end{eqnarray}
and
\begin{eqnarray}
\gamma = 2\pi  p( \omega_{Sk} ) \lambda_k^2|_{ \omega_{Sk} +
\omega_{31} = \omega_P} \nonumber
\end{eqnarray}
is the parameter describing the rapidity of the exponential
evolution to the entangled atomic state, $p(\omega_k)$ denotes the
density of states corresponding to the Stokes field, and
\begin{eqnarray}
\Delta = -{\cal P} \left\{ \int_{- \infty}^{\infty}
\frac{p(\omega_{Sk}) \lambda_k^2 d \omega_k}{\omega_k +
\omega_{31} - \omega_P} \right\} \nonumber
\end{eqnarray}
is a small frequency shift ($\cal P$ denotes the principle value
of the integral). Thus
\begin{eqnarray}
| \Psi (t) \rangle = e^{- \gamma t/2}e^{-i( \omega_P - \Delta )t}|
\psi_0 \rangle \nonumber \\ - \sum_k \frac{i \lambda_k
\sqrt{2}}{\gamma /2-i ( \omega_{Sk} + \omega_{31} - \omega_P -
\Delta)} \times \nonumber
\\ \times (e^{-i( \omega_{Sk} + \omega_{31})t}-e^{- \gamma
t/2}e^{-i( \omega_P - \Delta )t})| \psi_k \rangle \nonumber
\end{eqnarray}
and the system evolves exponentially to the maximum entangled
atomic state (3). In fact, this is a durable maximum entangled
atomic state because the direct single-photon transition $3
\leftrightarrow 1$ is forbidden. The lifetime of this entangled
state is defined by the slow non-radiative processes only.

Let us stress that the two advantages of the above considered
three-level two-photon process in comparison with the previous
scheme$^{4,5}$ are on the one hand the durability of the entangled
state and on the other hand the simple monitoring of entanglement
via detection of Stokes photon. We reckon that the quantum
information processing in the system under consideration can be
arranged in the same way as in Ref. 10.

The above long-life atomic entanglement can be interpreted as the
long-distance entanglement as well within the following
experimental scheme. Assume that one of the atoms is trapped in
the cavity which supports a single-photon Fock state of the pump
mode. The second atom passes through the cavity as shown in Fig.
2. Time of the propagation of the atom through the cavity defined
by the velocity of the atom should be long enough to provide the
preparing of the entangled state (3) with high probability. The
creation of this state is signalized by detection of the Stokes
photon. Then, the measurement of the state of the moving atom at
any distance from the cavity uniquely determines the state of the
trapped atom.

Concerning the practical realization of the above discussed
scheme, we should stress that the observation of single-atom
Raman-type process in an optical cavity has been reported
recently$^{11}$. In this work, the $^{85}Rb$ atom was used. The
excited state $2$ corresponds to $5P_{3/2}$ level, while the
ground $1$ and intermediate $3$ states are the $5S_{1/2}$
hyperfine levels separated by frequency $\omega_{31}= 3 GHz$,
while Stokes field has the wavelength $\lambda_S = 780 nm$. In
this case, the lifetime of the state $| 3 \rangle$ is at least ten
times longer than that for the excited state $2$.

Let us stress that the obtained result can be generalized on the
multi-atom case in the same way as for the conventional
single-photon process in two-level atoms$^6$. The increase of the
number of atoms should lead to a speeding-up of the evolution to
the entangled atomic state because of the Dicke-type process
caused by the photon exchange between the atoms (see Ref. 12).

The authors would like to thank J.H. Eberly and P.L. Knight for
useful discussions.

\begin{figure}
\caption{Scheme of Raman-type process in an atom. Solid arrows
show the allowed transitions. Wavy lines show the pump and Stokes
photons, respectively.}
\end{figure}

\begin{figure}
\caption{Scheme of creation of a durable two-atom entanglement.
Atom 1 is trapped in a cavity, while atom 2 can pass through the
cavity. Wavy lines show  the cavity and leaking out Stokes
photons.}
\end{figure}

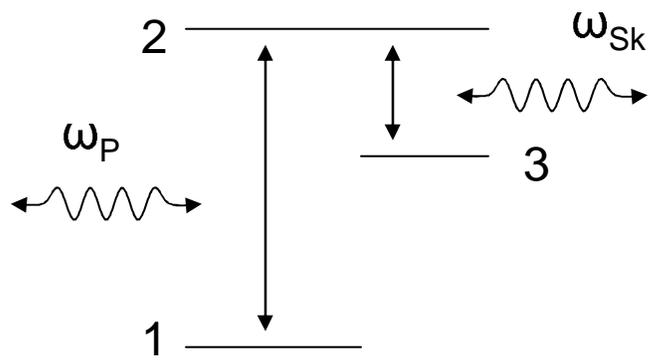
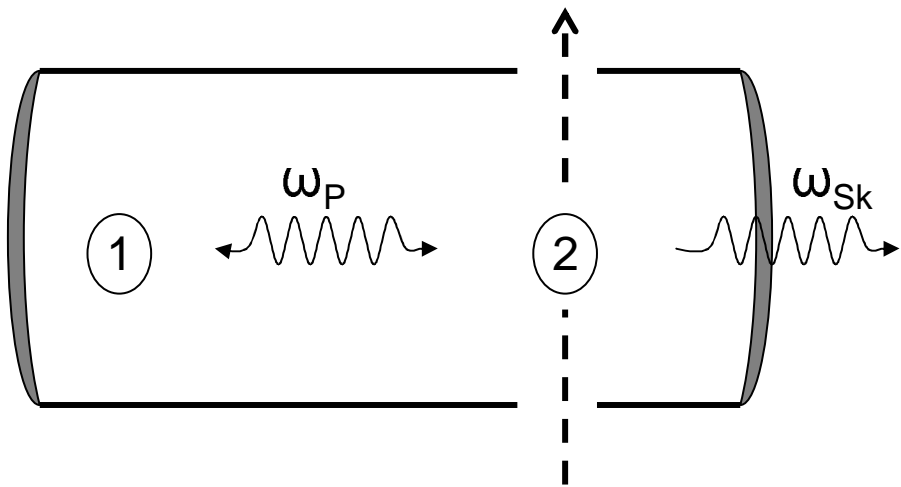
\end{document}